# Observation of Ultralong Valley Lifetime in WSe$_2$/MoS$_2$ Heterostructures

Jonghwan Kim[1]†, Chenhao Jin[1]†, Bin Chen[2], Hui Cai[2], Tao Zhao[1], Puiyee Lee[1], Salman Kahn[1], Kenji Watanabe[3], Takashi Taniguchi[3], Sefaattin Tongay[2], Michael F. Crommie[1,4,5], Feng Wang[1,4,5]*

[1] Department of Physics, University of California at Berkeley, Berkeley, California 94720, United States.

[2] School for Engineering of Matter, Transport and Energy, Arizona State University, Tempe, Arizona 85287, United States

[3] National Institute for Materials Science, 1-1 Namiki, Tsukuba, 305-0044, Japan.

[4] Material Science Division, Lawrence Berkeley National Laboratory, Berkeley, California 94720, United States.

[5] Kavli Energy NanoSciences Institute at University of California Berkeley and Lawrence Berkeley National Laboratory, Berkeley, California 94720, United States.

† These authors contributed equally to this work

* Correspondence to: fengwang76@berkeley.edu

**The valley degree of freedom in two-dimensional (2D) crystals recently emerged as a novel information carrier in addition to spin and charge[1,2]. The intrinsic valley lifetime in 2D transition metal dichalcoginides (TMD) is expected to be remarkably long due to the unique spin-valley locking behavior, where the inter-valley scattering of electron requires simultaneously a large momentum transfer to the opposite valley and a flip of the electron spin[1,2]. The experimentally observed valley lifetime in 2D TMDs, however, has been limited to tens of nanoseconds so far[3-6]. Here we report efficient generation of microsecond-long lived valley polarization in $WSe_2/MoS_2$ heterostructures by exploiting the ultrafast charge transfer processes in the heterostructure that efficiently creates resident holes in the $WSe_2$ layer. These valley-polarized holes exhibit near unity valley polarization and ultralong valley lifetime: we observe a valley-polarized hole population lifetime of over 1 μs, and a valley depolarization lifetime (i.e. inter-valley scattering lifetime) over 40 μs at 10 Kelvin. The near-perfect generation of valley-polarized holes in TMD heterostructures with ultralong valley lifetime, orders of magnitude longer than previous results, opens up new opportunities for novel valleytronics and spintronics applications.**

Atomically thin layers of semiconducting transition metal dichalcogenides (TMDs) exhibit unique electronic bandstructure[7,8] and fascinating physical properties[9,10]. A pair of degenerate direct bands are present at the K and K' points in the momentum space of hexagonal TMD monolayers, giving rise to a new valley degree of freedom known as the valley pseudospin[1,2]. The strong spin-orbital coupling present in TMDs further locks the valley pseudospin to specific electron and hole spins for electronic states close to the bandgap[1,2]. These coupled spin and valley degrees of freedom in TMDs can open up new ways to encode and process information for valleytronics; and they can be controlled flexibly through optical excitation, electrostatic gating, and heterostructure stacking. In particular, the spin-valley locking suggests that the intrinsic valley lifetime can be extremely long, because a change of valley pseudospin requires a rare event with a large momentum transfer (from K to K' valley) and an electron spin flip at the same time.

Tremendous progresses have been made in exploring the valley pseudospin of 2D TMDs, ranging from optical generation and detection of valley polarization[11-13], to manipulation of valley pseudospin state with optical and magnetic field[14-18], and to observation of valley Hall effect[19]. Many challenges, however, still exist for potential valleytronics applications. Chief among them is the relatively short valley lifetime. It was recently shown both theoretically and experimentally that the valley lifetime of excitons in TMD monolayers is severely constrained by the electron-hole exchange interaction through the Maialle-Silva-Sham mechanism[20-24], which can annihilate an exciton in one valley and create another exciton in the other valley, i.e. depolarize the valley pseudospin, within picoseconds. The valley pseudospin of individual electrons or holes, however, is not affected by this mechanism and can have much longer lifetime. Indeed, photo-induced valley polarization of resident carriers in TMD monolayers are reported to have much longer valley lifetime[3-5]. Bright interlayer exciton in type-II van der Waals heterostructure of TMDs provides another way to achieve longer valley lifetime, where electrons and holes are separated into different layers and the electron-hole exchange interaction is strongly suppressed. However, experimentally observed valley lifetime for either resident carriers in TMD monolayers or indirect excitons in TMD heterostructures has been limited to few tens of nanosecond so far[3-6]. Here we report efficient generation of ultralong lived valley polarization in $WSe_2/MoS_2$ heterostructures. Using ultrafast pump-probe spectroscopy that covers time scale from femtoseconds to microseconds, we show that perfectly valley-polarized holes can be generated in the $WSe_2$ layer within 50 fs owing to the ultrafast charge transfer processes in the $WSe_2/MoS_2$ heterostructure[25,26]. These valley-polarized holes exhibit a population decay lifetime of over 1 μs, and a depolarization lifetime (i.e. inter-valley scattering lifetime) over 40 μs at 10 Kelvin, which is orders of magnitude larger than previously reported values. The near unity valley polarization and ultralong valley lifetime observed here will enable new ways to probe and manipulate valley and spin degrees of freedom in TMDs.

We investigate high quality $WSe_2/MoS_2$ heterostructures using polarization-resolved pump-probe spectroscopy. Fig. 1b shows the optical microscopy image of a representative $WSe_2/MoS_2$ heterostructure. The $WSe_2$ (encircled by the blue dashed line) and $MoS_2$ (encircled by the red dashed line) monolayers are first exfoliated mechanically from bulk crystals onto $SiO_2$/Si substrates, and then stacked to form the heterostructure (denoted by the black dashed line) by a dry transfer method using a polyethylene terephthalate (PET) stamp (See Methods). The heterostructure region can be visualized most strikingly in the photoluminescence (PL) image (Fig. 1c), where the PL at the $WSe_2$ A-exciton resonance (1.65 eV at room temperature) is quenched by more than 4 orders of magnitude in the heterostructure region compared to the $WSe_2$-only region. This quenching of photoluminescence is a signature of the type-II heterojunction in $WSe_2/MoS_2$ heterostructures, where the conduction band minimum and the valence band maximum reside in the $MoS_2$ and $WSe_2$ layers, respectively, and an ultrafast charge transfer process takes electron from the $WSe_2$ to $MoS_2$ layer upon photoexcitation of $WSe_2$ excitons (Fig. 1a).

The ultrafast electron transfer process in the heterostructure allows for efficient generation of valley-polarized holes in $WSe_2$, as illustrated in Fig. 2a: Resonant photoexcitation with left circularly-polarized (LCP) light selectively creates electron-hole pairs (i.e. excitons) at the K valley in the $WSe_2$ layer. After the excitation, the electrons can transfer to the MoS2 layer within 50 fs and leave behind resident holes in the K valley of the WSe2 layer. These valley polarized holes can exhibit ultralong lifetime: The population and valley polarization relaxation processes of resident holes, such as radiative recombination and exchange interaction (Maialle-Silva-Sham mechanism), are dramatically suppressed because the holes are well-separated from electrons in not only real space but also momentum space. Compared with the previously studied valley polarization of bright interlayer exciton (which requires the presence of both electrons and holes), the lifetime of resident holes is not limited by the decay of electrons in the other layer. To achieve the longest valley lifetime, we choose mechanically exfoliated and stacked $WSe_2/MoS_2$ heterostructures: the exfoliated TMD layers exhibit much higher quality and fewer defects compared with chemical vapor deposition grown samples; and the $WSe_2/MoS_2$ heterostructure features the largest band offset among all TMD combinations[27,28] so that electrons and holes are well confined in separate layers.

We investigate the dynamic evolution of valley-polarized holes in the $WSe_2$ layer using polarization-resolved pump-probe spectroscopy. The LCP pump pulses generate valley-polarized holes in the $WSe_2$ layer. Such valley imbalance leads to a difference in optical absorption of the heterostructure for LCP and RCP light close to the $WSe_2$ A-exciton resonance, and can thereby be probed by pump-induced changes in the reflection contrast (RC) spectra of circularly polarized probe pulses. The dynamic evolution of the polarization-resolved ΔRC from femtosecond to microsecond was measured by combining a mechanical delay line (from femtoseconds to nanoseconds) and an electronic delay (from nanoseconds to microseconds).

The upper panel of Fig. 2b displays a photo-induced circular dichroism (CD) spectrum of the heterostructure probed at 3 ns after the pump excitation, and the bottom panel of Fig. 2b shows the RC spectrum of the heterostructure. The RC spectrum is dominated by the optical absorption near the WSe$_2$ A-exciton resonance peak at 1.72 eV[29,30]. In the pump-probe study, we choose an excitation energy at 1.78 eV (blue arrow in Fig. 2b) to selectively excite WSe$_2$ but not MoS$_2$ (which has an optical bandgap of 1.92 eV). The CD signal is measured through $-\Delta(RC_{\sigma+}-RC_{\sigma-})$, the difference between the photo-induced changes in the reflection contrast of LCP ($RC_{\sigma+}$) and RCP ($RC_{\sigma-}$) light. The CD spectrum exhibits a prominent resonant feature around the A-exciton transition, and its magnitude is directly proportional to valley-polarized hole density, i.e. the difference between the hole density in the K valley ($p_+$) and in the K' valley ($p_-$).

Fig. 2c shows the time evolution of the CD signal with 1.71 eV probe photons at 10 Kelvin. The dynamic response from femtoseconds to nanoseconds is measured using a mechanical delay line, and it shows an almost constant CD signal from 300 fs to 3.5 ns (black curve in the inset). To capture the longer-term dynamics, we use an RF-coupled diode laser synchronized to the femtosecond pulses to generate 3 ns long probe pulses with electronically defined time delay up to a few microseconds (see Methods). The CD signal from nanosecond to microsecond time scale is shown as red squares in main panel. Strikingly, the CD signal remains appreciable even after several microseconds, and the slowest decay component shows a lifetime of more than 1 μs. (The relatively fast initial decay is largely due to interactions between photo-excited carriers. See pump fluence dependence in Supplementary Information Part1). This microsecond lifetime of valley-polarized hole density is orders of magnitude longer than those reported previously[3-6].

To separate the contributions from the population decay and the inter-valley scattering to the overall lifetime of valley-polarized holes, and to quantify the degree of valley polarization upon optical initialization, we examine in more detail the dynamic behavior and spectra dependence of the photo-induced valley polarization in $WSe_2/MoS_2$ heterostructures.

One key figure of merit in valley initialization and control is the degree of valley polarization, as defined by:

$$P = \frac{p_+ - p_-}{p_+ + p_-}$$

where $p_+$ and $p_-$ are hole densities in K and K' valleys, respectively. Here we need to establish a quantitative relation between the hole density in a specific valley and the corresponding circularly polarized ΔRC spectra. Towards this goal, we analyze the photo-induced absorption changes in individual K and K' valleys by examining the ΔRC spectra for LCP and RCP light separately. Fig. 3a and 3b show the ΔRC spectra at K and K' valley, respectively, at 3ns after resident holes are created in the K-valley of $WSe_2$ through LCP light excitation. Distinctively different absorption changes are observed for the K ($\Delta RC_{\sigma+}$, black squares in Fig. 3a) and K' valleys ($\Delta RC_{\sigma-}$, red circles in Fig. 3b): The holes present in the K-valley can modify further absorption at the K valley through a combination of the phase-space filling and Burstein-Moss effects, as previously observed for exciton states in quantum wells[31,32]. It leads to a reduction of the exciton absorption oscillator strength accompanied by a slight blueshift of the exciton resonance, as illustrated in the inset of Fig. 3a. Consequently, $\Delta RC_{\sigma+}$ for the K valley is dominated by an overall absorption reduction with a small absorption increase at the higher energy side. The effect of K-valley holes on the absorption of the K' valley, however, is completely different because there is no Pauli blocking from phase-space filling. Instead, $\Delta RC_{\sigma-}$ is characterized by a decrease in the exciton absorption and an increase in the trion absorption with the total absorption oscillator strength conserved (inset in Fig. 3b).

The understanding that holes present in one valley will reduce the overall optical absorption oscillator strength at the same valley but not the opposite valley allows for a direct experimental determination of hole population in each valley: $p_+$ and $p_-$ are proportional to the integration of the $\Delta RC$ signal over frequency for LCP and RCP light, respectively. Following this approach, we find that an almost prefect valley polarization was created by the LCP excitation pulses: holes exist only in the K valley but not the K' valley of $WSe_2$ within the experimental uncertainty (Fig. 3c). This efficient generation of valley polarized holes presumably benefits from the ultrafast electron transfer process in the heterostructure, which separates the electrons and holes even before the valley depolarization of excitons has a noticeable effect.

Next we investigate quantitatively the time evolution of valley-polarized holes. The decay of CD signal, proportional to the valley-polarized hole density (i.e. $p_+ - p_-$), has two different contributions: (1) a population decay of the total hole density (i.e. $p_+ + p_-$) and (2) inter-valley scattering that reduces the degree of valley polarization P=$(p_+ - p_-)/(p_+ + p_-)$. These two contributions can be obtained separately by examining the time evolution of $\Delta RC_{\sigma+} + \Delta RC_{\sigma-}$ and $(\Delta RC_{\sigma+} - \Delta RC_{\sigma-})/(\Delta RC_{\sigma+} + \Delta RC_{\sigma-})$, respectively.

Fig. 4a and 4b show the photo-induced difference ($\Delta RC_{\sigma+} - \Delta RC_{\sigma-}$) and sum ($\Delta RC_{\sigma+} + \Delta RC_{\sigma-}$) signals, respectively, at 3 ns, 80 ns and 550 ns pump-probe delay. All the spectra are normalized to one. Both the difference and the sum spectra exhibit a constant profile around the A-exciton resonance (in the spectral range of 1.69-1.78 eV). In the sum response (Fig. 4b), a weak signal appears around 1.66 eV over time, presumably due to some defect states that decay slowly. These defect states, however, do not distinguish the K and K' valley and do not show any effect in the difference response (Fig. 4a). Because both the difference and sum spectra remain constant profiles within 1.69-1.78 eV, we can use a single probe photon energy at 1.71 eV to obtain the time evolution of the difference and sum signals, which characterize the difference and sum of hole densities in the K and K' valley, respectively.

Fig. 4c displays the normalized decay dynamics of the total ($p_+ + p_-$, black dots) and valley-polarized ($p_+ - p_-$, red squares) hole densities in the heterostructure, as well as the degree of valley polarization (P, blue triangles). We found that the decay of valley-polarized holes ($p_+ - p_-$) is very similar to that of the total hole density ($p_+ + p_-$), indicating that the 1 µs decay lifetime observed in CD signals is dominated by a population decay of holes. The valley depolarization lifetime from inter-valley scattering, on the other hand, is much longer: the degree of valley polarization does not show any apparent decay in 2.5 µs (except a small decrease within the first hundred nanoseconds). A conservative lower limit of the depolarization lifetime is 40 µs (see Supplementary Information). However, there is significant uncertainty because of the minimal decay, and the upper limit can be many hundreds of µs.

At elevated temperatures, the valley depolarization becomes faster. Fig. 4d shows the temperature dependence of the valley polarization decay. All decay curves show a weak fast component and a dominant slow component. We fit the measured decay dynamics (symbols in Fig. 4d) with a bi-exponential decay (solid lines), and focus on the behavior of the slow decay component. The slow depolarization lifetime changes from 10 ns at 77 K to above 40 µs at 10 K, as shown in Fig. 4e. This strong temperature dependence of valley depolarization lifetime suggests an energy-activated mechanism for inter-valley hole scattering in WSe$_2$/MoS$_2$ heterostructures. However, a microscopic understanding of the inter-valley scattering processes in the heterostructure will require in depth theoretical investigations, and is beyond the present work.

Our studies show that an almost perfect valley polarization of holes can be generated optically in WSe$_2$/MoS$_2$ heterostructures. The valley-polarized holes exhibit a population decay lifetime of ~ 1 µs and a depolarization lifetime approaching 100 µs at 10K. Presumably the population lifetime can be improved with better sample quality to reduce defect traps and with optimized heterostructure design to better separate electrons and holes. Therefore a valley-polarized hole population with 100 µs should be possible. These valley-polarized holes are also spin polarized due to the spin-valley locking in WSe$_2$. Such long-lived valley and spin polarizations can open up new opportunities for valleytronics and spintronics applications based on two-dimensional van der Waals heterostructures.

**Methods:**

**Heterostructure preparation**

WSe$_2$/MoS$_2$ heterostructures were prepared with polyethylene terephthalate (PET) stamp by dry transfer method[33]. Monolayer WSe$_2$, MoS$_2$, and hBN flakes were first exfoliated onto silicon substrate with 90 nm oxide layer. We used PET stamp to pick up the hBN flake and monolayer MoS$_2$ in sequence. The PET stamp with hBN/MoS$_2$ was then stamped onto a monolayer WSe$_2$ flake to form a MoS$_2$/WSe$_2$ heterostructure. Polymer and samples were heated up at 60 ℃ in the pick-up and at 130 ℃ in the stamp process. Finally we dissolved the PET in dichloromethane for 12 hours at room temperature.

**Generation of a nanosecond optical pulse with electronically controlled time delay**

To generate optical pulses with electronically defined time delay, we first generated electronic pulses synchronized to the femtosecond laser. The femtosecond laser output (Pharos, Light conversion) has a repetition rate of 150 kHz defined by a regenerative amplifier, whose internal clock is used to trigger an electronic pulse generator (HP 8082A). The output electronic pulses has ~ 3 nanosecond pulse duration, which is then converted to optical pulses by a RF-coupled laser diode module (Thorlabs TCLDM9). The relative time delay between the femtosecond laser output and diode laser output can thereby be accurately controlled with the electronic delay in the pulse generator.

**Figures:**

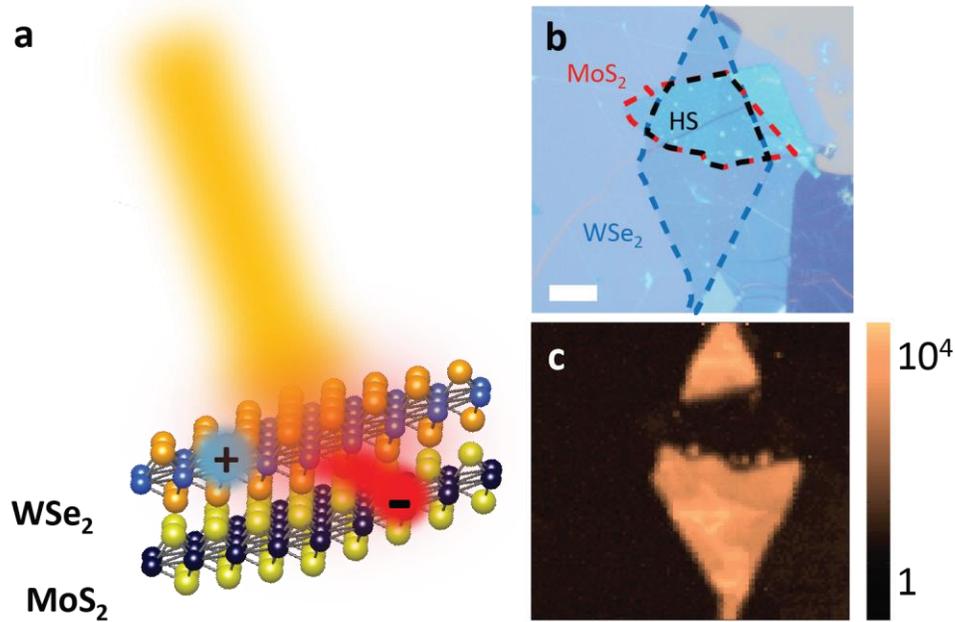

**Figure 1 | Ultrafast charge transfer process in the WSe$_2$/MoS$_2$ heterostructure. a,** Illustration of ultrafast electron transfer process in WSe$_2$/MoS$_2$ heterostructure. WSe$_2$/MoS$_2$ heterostructure forms a type II heterojunction where conduction band minimum and valence band maximum reside in MoS$_2$ and WSe$_2$, respectively. Photoexcited electrons transfer rapidly to MoS$_2$ layer whereas holes remain in WSe$_2$ layer. **b,** Optical microscope image of a representative WSe$_2$/MoS$_2$ heterostructure. Blue, red and black dashed lines encircle WSe$_2$, MoS$_2$ and heterostructure (HS) regions, respectively. Scale bar corresponds to 10 μm. **c,** Photoluminescence (PL) image of the WSe$_2$/MoS$_2$ heterostructure at WSe$_2$ A-exciton resonance (1.65 eV) at room temperature. PL is quenched by 4 orders of magnitude in the heterostructure compared to the WSe$_2$-only region due to the ultrafast electron transfer process in the heterostructure.

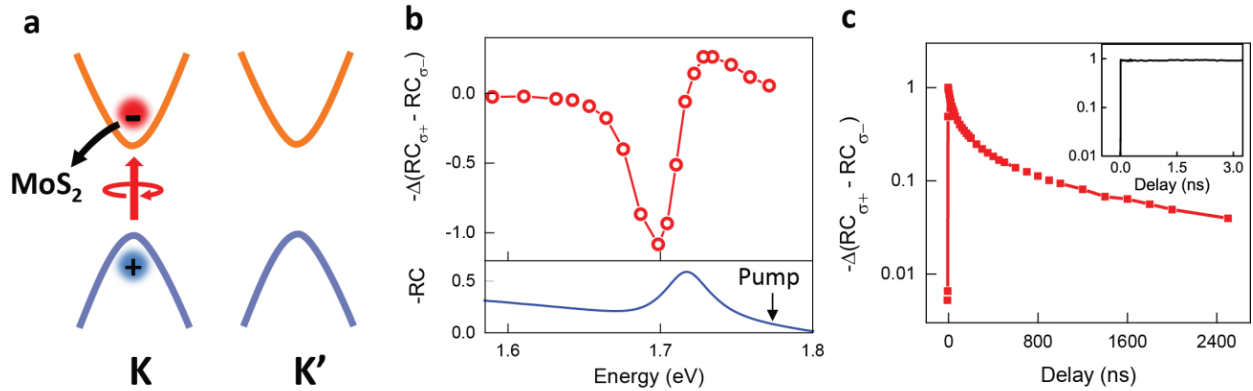

**Figure 2 | Photo-induced circular dichroism (CD) signal of the WSe$_2$/MoS$_2$ heterostructure at 10 K. a,** Schematic of valley-polarized hole generation in WSe$_2$ layer within the heterostructure. Upon photoexcitation with left circularly-polarized light, excitons are resonantly created in K valley of WSe$_2$ layer. Ultrafast charge transfer process then efficiently transfers electrons to MoS$_2$ layer and leaves resident holes at K valley in WSe$_2$ layer. **b,** (Top panel) Photo-induced CD spectrum at 3ns and (Bottom panel) reflection contrast (RC) spectrum of the WSe$_2$/MoS$_2$ heterostructure. RC spectrum is dominated by optical absorption near WSe$_2$ A-exciton resonance at 1.72 eV. The CD spectrum, $-\Delta(RC_{\sigma+} - RC_{\sigma-})$, shows prominent resonant feature near WSe$_2$ A-exciton peak under LCP pump light at 1.78 eV (black arrow in RC spectrum). **c,** Decay dynamics of the resonant CD signal at 10 K. No decay is observed within 3.5 ns (Inset). The decay curve over a longer time scale shows a significant slow decay component with a lifetime more than 1 μs.

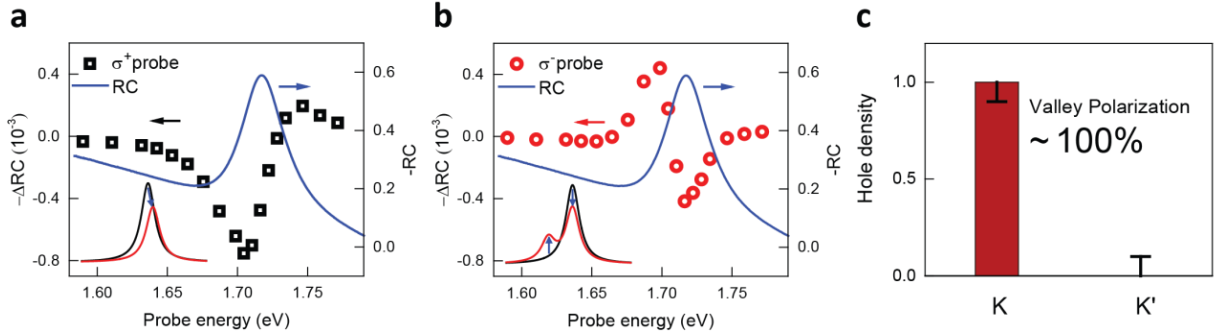

**Figure 3│An almost perfect valley polarization.** Resident holes in K-valley induce distinctively different absorption changes for the K **(a)** and K' **(b)** valleys; reflection contrast of the heterostructure (blue solid line) is also shown for comparison. **a,** The absorption change in the K-valley features an overall reduction of the absorption oscillator strength and a slight blueshift of the exciton resonance, as illustrated in the inset. This spectral change can be understood by the phase-space filling and Burstein-Moss effects. **b,** The absorption change in the K' valley shows a transfer of oscillator strength from exciton to trion absorption due to formation of intervalley trions. However, the total oscillator strength is unaffected since there is no Pauli-blocking effect. **c,** The density of resident holes in K and K' valley obtained by the integrated oscillator strength change of LCP and RCP light, which suggests near-perfect valley polarization (within 10% experimental uncertainty). The total hole density is normalized to 1.

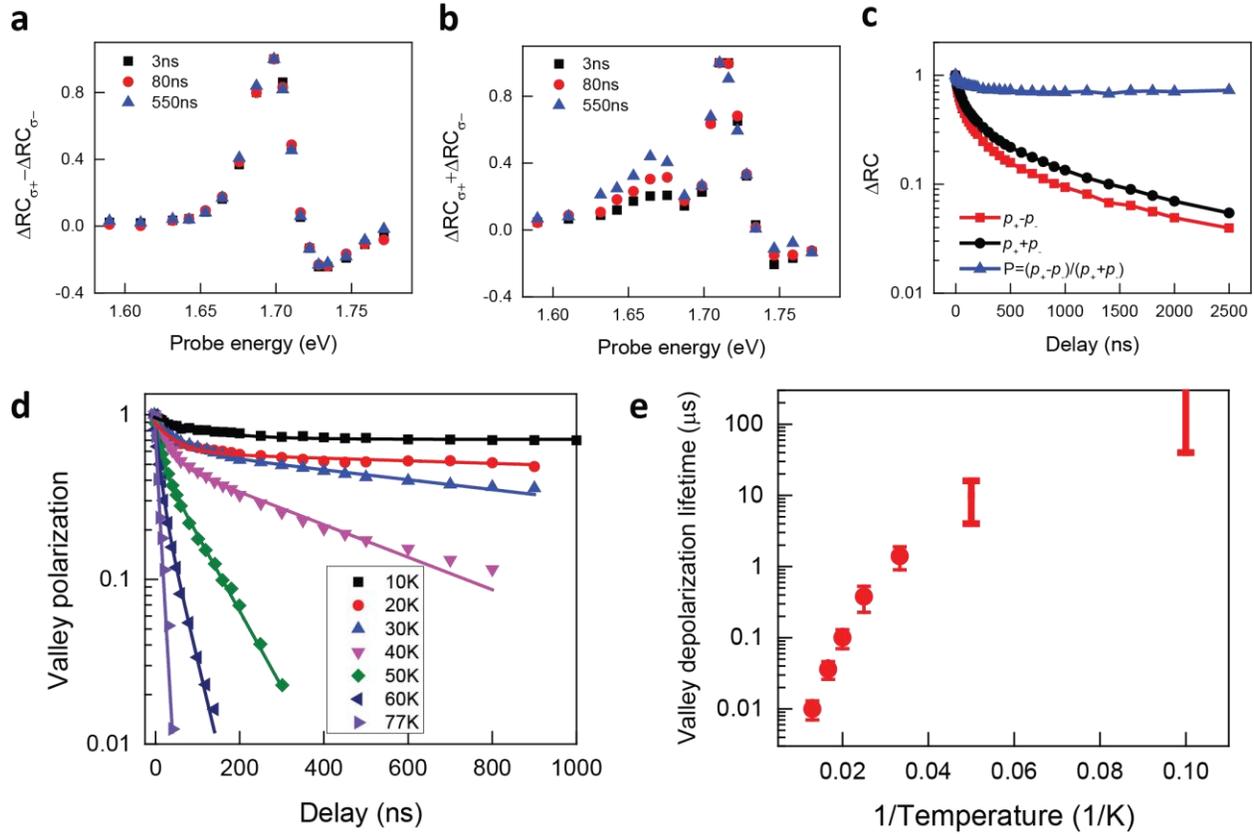

**Figure 4 | Ultralong valley depolarization lifetime. a and b,** Photo-induced difference **(a)** and sum **(b)** responses of the two valleys at a pump-probe delay of 3ns (black), 80ns (red) and 550ns (blue). All spectra were measured at 10 Kelvin and normalized to 1. Both the difference and sum spectra show constant profile over time, except for a weak signal around 1.66 eV in the sum response due to some defect states that decay slowly. **c,** Decay dynamics of the total hole population $p_+ + p_-$ (black dots), valley-polarized hole population $p_+ - p_-$ (red squares), and degree of valley polarization P (blue triangles) obtained with a probe energy of 1.71 eV. The decay of the valley-polarized hole population of ~ 1 μs is mainly due to the total population decay. The valley polarization, however, does not show any apparent decay in 2.5 μs, corresponding to an ultralong valley depolarization lifetime approaching 100 μs. **d,** Temperature-dependent decay dynamics of valley polarization from 10 K to 77 K (symbols). Solid lines are bi-exponential decay fitting of experimentally measured decay dynamics, with decay lifetime of dominant slow components summarized in **e**. The valley depolarization lifetime changes strongly with the temperature, suggesting an energy-activated mechanism in intervalley hole scattering.

# Supplementary Materials for

# Observation of Ultra-long Valley Lifetime in WSe$_2$/MoS$_2$ Heterostructures

Jonghwan Kim[1]†, Chenhao Jin[1]†, Bin Chen[2], Hui Cai[2], Tao Zhao[1], Puiyee Lee[1], Salman Kahn[1], Kenji Watanabe[3], Takashi Taniguchi[3], Sefaattin Tongay[2], Michael F. Crommie[1,4,5], Feng Wang[1,4,5]∗

[1] Department of Physics, University of California at Berkeley, Berkeley, California 94720, United States.

[2] School for Engineering of Matter, Transport and Energy, Arizona State University, Tempe, Arizona 85287, United States

[3] National Institute for Materials Science, 1-1 Namiki, Tsukuba, 305-0044, Japan.

[4] Material Science Division, Lawrence Berkeley National Laboratory, Berkeley, California 94720, United States.

[5] Kavli Energy NanoSciences Institute at University of California Berkeley and Lawrence Berkeley National Laboratory, Berkeley, California 94720, United States.

† These authors contributed equally to this work

∗ Correspondence to: fengwang76@berkeley.edu

**1. Pump fluence dependence of the photo-induced circular dichroism signal in the heterostructure**

Fig. S1 shows the pump-induced circular dichroism (CD) signal of WSe$_2$/MoS$_2$ heterostructure measured through $-\Delta(RC_{\sigma+} - RC_{\sigma-})$ with different pump fluence from 65 nJ·cm$^{-2}$ to 650 nJ·cm$^{-2}$. The pump light energy is at 1.78 eV and probe light at 1.71 eV. All spectra are normalized by pump fluence. The fluence-normalized CD signal shows a constant initial amplitude, indicating that the population of valley-polarized holes generated in WSe$_2$ is linearly proportional to the pump fluence. This is consistent with the efficient generation of valley-polarized holes due to the ultrafast charge transfer process in the heterostructure. On the other hand, the decay dynamics of the CD signal depends sensitively on pump fluence: the initial fast decay component decreases with lower pump fluence. Such fluence dependence suggests that higher order processes (such as Auger recombination) that involves interactions between photo-generated carriers dominate the initial decay dynamics when the hole density is high. To avoid this complication, we focus on the slow decay component characterizing single-hole behaviour at low excitation density.

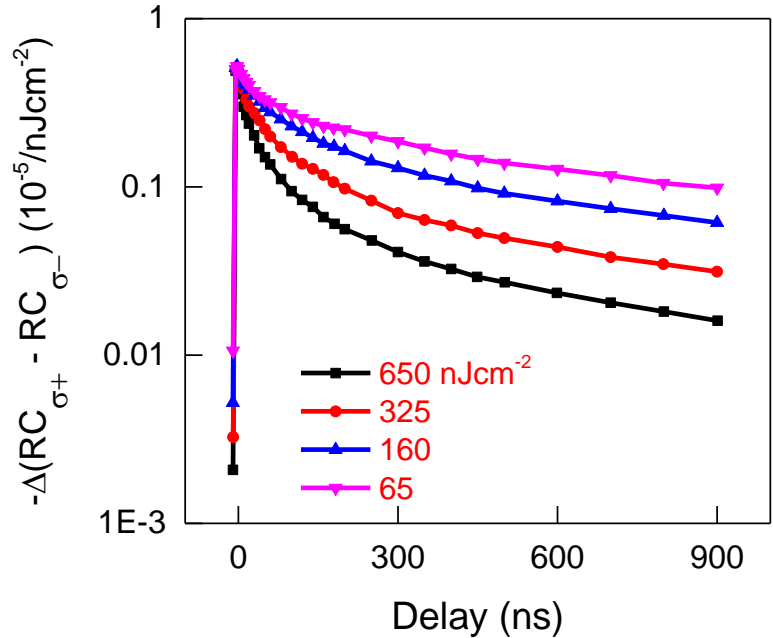

**Fig. S1** CD signal of the heterostructure at 10K with different pump fluence from 65 nJ·cm$^{-2}$ to 650 nJ·cm$^{-2}$. The initial fast decay components depend sensitively on the pump fluence, which is dominated by interactions between photo-excited carriers.

## 2. Estimation of valley depolarization lifetime at 10 Kelvin

The degree of valley polarization remains almost a constant up to 2.5 μs at 10 Kelvin (Fig. 4c), indicating an ultralong valley depolarization lifetime. We estimate the lower limit of valley depolarization lifetime based on our experimental uncertainty. We determine the experimental uncertainty of valley polarization by calculating the standard deviation $\sigma_P$ of the measured data between 500ns and 2500ns (see Fig. 4C in the text), and obtain a mean value $\bar{P} = 0.71$ and $\sigma_P = 0.012$. As a conservative estimation, the experimental uncertainty is within $3\sigma_P = 0.036$, which indicates that the valley polarization decay in 2μs is within 5%. This uncertainty gives a lower limit of depolarization lifetime to be 40 μs, and there is no upper bound. The actual value for the valley depolarization lifetime can well be hundreds of μs at 10K.